# On the unexpected fate of scientific ideas
# An archeology of the Carroll group

Jean-Marc Lévy-Leblond
Université de Nice, France
jmll@unice.fr

> …notwithstanding the sagacious advice
> by Lewis Carroll himself, who wrote:
> «It's no use going back to yesterday,
> because I was a different person then.» [1]

## Abstract

In 1965, I published a paper, exhibiting a hitherto unknown limit of the Lorentz group, which I christened "Carroll group" because of its seemingly paradoxical physical contents. Since I saw it as more curious than relevant, I published it in French in a journal somewhat afar from the mainstream of theoretical physics at that time. It was most gratifying to witness the quite unexpected favour this paper started to enjoy half a century later, so much that a so-called "Carrollian physics" is now developing, with applications in various domains of forefront theoretical physics, such as quantum gravitation, supersymmetry, string theory, etc. I offer this narrative as an example of the very diverse time scales with which scientific ideas may develop — or not.

## 1 Paying tributes

Let me start by diving in a more remote time than my own in order to acknowledge my debt to some of our predecessors who stressed the importance of group structure as one of the pillars of theoretical physics.

A prominent figure in my personal Pantheon is Paul Langevin (1872-1946), whose 150[th] birthday we recently celebrated. Not only did he contribute to crucial developments of Einsteinian relativity, but in his most resolute and successful endeavour for clarifying and popularizing it, he insisted as early as 1911, on the group theoretical perspective, writing

> «It is an experimental fact that the equations between physical quantities by which we translate the laws of the outside world, must have exactly the same form for different groups of observers, for various systems of reference in uniform translation relative to each other. This requires, in the language of mathematics, that these equations admit of a *group of transformations* corresponding to a change of reference system to another moving relative to it. The equations of physics must be preserved for all transformations of this group. In such a transformation, when one moves from one reference system to another, measures of various magnitudes, especially those that are related to space and time, are changed in a manner that corresponds to the structure of these notions.» [2]

It is worthwhile recalling here what Einstein, who maintained a lifelong friendly relationship with Langevin (Figure 1), wrote in his funeral tribute:

> «It appears to me as a foregone conclusion that he would have developed the special relativity theory, had not that been done elsewhere.» [3]

One may well admire the elegance of the last words…

On a more private level, I wish to salute the memory of my group-theoretical masters Louis Michel (1923-1999) and François Lurçat (1927-2012), as well as of my friend and collaborator Henri Bacry (1928-2010).

## 2 Siring the Carroll group

After my PhD, dedicated to the Galilei group and its representations in the line of Wigner's epochal paper on the Lorentz group [4], I soon started to teach Einsteinian relativity, stressing both its analogies and its differences with Galilean relativity. While preparing my lectures, I stumbled upon an unexpected difficulty, concerning the validity of the Galilean approximation to the Lorentz transformations. Choosing the natural system of units where the limit velocity $c$ is taken as unity, the Lorentz transformations relating spacetime intervals in two equivalent reference frames with relative velocity $v$ take the form:

$$\begin{cases} \Delta x' = \gamma(\Delta x - v\Delta t) \\ \Delta t' = \gamma(\Delta t - v\Delta x) \end{cases} \quad \text{where} \quad \gamma = \sqrt{1 - v^2}.$$

If we now wish to obtain approximate formulas in the situation where $v \ll 1$, the Galilean group law transformations

$$\begin{cases} \Delta x' = \Delta x - v\Delta t \\ \Delta t' = \Delta t \end{cases}$$

do not arise obviously unless we require the additional condition $\Delta x \ll \Delta t$. It thus appears that the validity of the Galilean approximation requires not only velocities small with respect to the limit velocity, but also large time-like intervals, a condition which has stood implicit in most (not to say all) treatments of the subject.

It is then natural to inquire what happens in the opposite case of large space-like intervals, that is for $\Delta t \ll \Delta x$, yielding the transformation laws

$$\begin{cases} \Delta x' = \Delta x \\ \Delta t' = \Delta t - v\Delta x \end{cases}.$$

These transformations obviously form a group, which, as well as the Galilei group, is a contraction of the Lorentz group [5] and thus seemed as well worth of recognition [6]. In a world governed by such an invariance group, causality almost completely disappears, since time-ordering is only preserved along the timelines at given space points. For this reason, I ventured to propose the name "Carroll group" for this alternate degenerate limit of the Lorentz group [7].

An illustrative way of expressing the situation is that, while the Galilei group appears as the limit of the Lorentz group when a rescaling of Minkowski space flattens the light cone on the constant time hyperplane, the Carroll group emerges when the light cone closes up on the time axis (Figure 2). This leads to another manner of considering the Carroll group, by recovering dimensional velocities. Since the limit velocity $c$ is but the slope of light rays in Minkowski spacetime, it results that, whereas the Galilei group corresponds to the well-known limit $c \to \infty$, the Carroll group may be seen as resulting from the inverse limit, that is $c \to 0$, weird as this limit may appear at first glance.

At the time, I was convinced that, because of the acausal nature of a universe obeying Carrollian invariance, the usefulness of the Carroll group was very low, and I apologized for begetting it, arguing that my purpose was mainly pedagogical. So little did I believe in its fate that I published my result in French, in a journal which did not belong to the mainstream publications of theoretical physics, concluding my paper by stating with some cheekiness that «theoretical physics has recently shown itself to be friendly enough for many groups with a limited physical interest; this is why I have not too much scruple in bringing to light this degenerate cousin of the Poincaré group.».

It took me a long time to learn that a paper quite similar to mine had been written independently by an Indian colleague, N. D. Sen Gupta, working at the Bombay Tata Institute for Fundamental Research [8]. Rather unexpectedly, his paper had been published very soon after mine, and, as far as I know, might even had been written before. While it appeared in a journal at that time more prominent in theoretical physics, the long delay in my recognition of Sen Gupta's work and the paucity of any references to it in the literature for many years bear witness to the little interest elicited by our almost simultaneous small discovery.

## 3 A terminological excursus

*"Relativity" ?*

With the benefit of half a century of personal and collective maturation, I wish to indulge here in a piece of self-criticism concerning the use of the term "non-relativistic" in the title and text of my old paper. The word was routinely used until late in the XXth century to mean, as many dictionaries still propose, "not based on or not involving the special relativity of Einstein". However, it has become clear today that Einsteinian relativity is not the only consistent one and that the kinematical structure of classical mechanics obeys a relativity theory of its own, now known as Galilean relativity. One might object to this denomination that it is somewhat anachronistic, as the general notion of space-time symmetries would take almost three centuries after Galileo to emerge. Nonetheless, a most important paragraph in Galileo's seminal *Dialogo* shows quite clearly that he had fully understood the invariance of physical laws with respect to changes of reference frames with uniform velocity [9].

But we might well take one more step in the revision of the current terminology. Indeed, the term "relativity" itself, which, concerning its use in physics, dates back to Poincaré [10] , may be considered as a misnomer. As early as 1948, an undisputable authority, namely A. Sommerfeld, wrote:

> «[The theory of space-time] is an *Invariantentheorie* of the Lorentz group: the relativity of space and time is not the essential thing, which is the independence of the laws of nature from the point of view of the observer.» [11]

This independence/invariance is characterized by the intrinsic structure of spacetime, that is, what I believe natural to name a **chronogeometry**, exactly as we call elementary geometry the theoretical structure of Euclidean space, with the term "geometry" having been generalised to describe the structure of variously defined spaces, according to Klein's Erlangen program [12].

*A proposal :*
— replace "relativity" by "chronogeometry"
— replace "non-relativistic" by "Galilean" (…or "Carrollian")

*"Speed-of-light"?*

Einstein's derivation of Lorentz transformations (1905) was based on the so-called second postulate, that of the invariance of light velocity. But Einsteinian chronogeometry is not intrinsically linked to the properties of light: indeed, as a universal structure of Minkowskian spacetime, it rules as well non-electromagnetic phenomena, such as strong interactions. As a matter of fact, it was realized as early as 1911 that the Lorentz group may be constructed without any appeal to the second postulate (Ignatowsky, Frank & Rothe), as many authors have rediscovered since [13]. Furthermore, suppose the photon has a non-zero mass, however small, a case which, obviously, cannot be excluded; then light would not travel with a non-invariant velocity…

> *A proposal :*
> — replace "speed of light" by "speed limit" or better: "Einstein constant".

### *"Group contraction"*

The notion of group contraction, which formalize the limiting process leading from Einsteinian chronogeometry to the Galilean (or Carrollian) one, was introduced just a century ago by Inönu and Wigner in the following terms:

> «We shall call the operation of obtaining a new group by a singular transformation of the infinitesimal elements of the old group a contraction of the latter. The reason for this term will become clear below. (…) In the limit $e = 0$ (if such a limit exists), one will have contracted the whole group to an infinitesimally small neighborhood of the group.»

The structure of the But this is hardly a proper description. In fact, the process goes rather the other way, that is, extending the structure of an infinitesimal neighbourhood of the group to that of a fully fledged new group. More than a simple extension, this change of structure in fact deserves to be considered as a distension.

> *A proposal:*
> — replace "group contraction" by "group distension"

Such discussions about the terminology of physics are by many considered as futile nitpicking: why, do they ask, should we care about words since we have the formulas to rely on? This is not the place to develop a detailed answer [14]. Let me only state that paying attention to our linguistic choices and assessing their relevance, may be of great significance for research, as it should go beyond formalism, for teaching, as it should go beyond technicalities, for popularisation, as it should go beyond catchwords.

## 4 The late blooming of Carrollian physics

As I had suspected right from the beginning, the decades following the appearance of the Carroll group on the theoretical scene, viewed very little references to my paper or Sen Gupta's.

From 1965 to the early 2000s, there were one or two quotations per year in journals as varied as *J. Math.Phys., Ann. Der Phys., J. of Phys. Comm., Nuovo Cimento, Int. J. Theor. Phys., Bull. Acad. R. Belg., Phys. Lett., J. Geometry & Phys.,* etc., dealing mainly with general considerations about abstract Group theory, Special relativity, Electromagnetism. But, to my utmost surprise, from the 2000s onwards, with a notable acceleration after 2010, more and more papers appeared dealing with Carrollian chronogeometry, with a concentration in *J. Math. Phys., J. of High Energy Phys., Class. & Qu. Gravity, Phys. Rev. D, Phys. Rev. Lett., General Relat. & Gravit., J. of Cosmology & Astroparticle, Phys. Lett.* (Figure 3). These works are now mainly concerned with General relativity, Field Theory, Gravitation, etc. Some keywords characterizing them are: Conformal structures, Asymptotic flat spacetimes, Nonrelativistic SUSY, Symplectic spacetime, Non-Riemannian isometries, Bondi-Metzner-Sachs group, Null manifolds, Cartan geometry, Anti-deSitter symmetry, BMS field theories, Tachyon cosmology, Flat holography, Chern-Simons supergravity, etc. [15]

While this renewal of interest may at first seem puzzling, the reason in fact is rather easy to understand. Indeed, if Galilean chronogeometry yields a simple approximate way to explore the portion of Minkowski spacetime interior to the lightcone, Carrollian chronogeometry furnishes a similar simple approximate way to explore the portion of Minkowski spacetime exterior to the lightcone. Even though the latter is an acausal region, it is a constitutive portion of spacetime and plays a significant role in many physical phenomena. This argument may be strengthened by considering how Lorentz transformations may be generated by a combination of Galilean and Carrollian ones [16].

A few years after the appearance of the Carroll group, it was shown by Henri Bacry and myself to take place in an overall description and classification of logically and physically possible chronogeometries [17], as summed up by an elegant diagram (Figure 4). This paper had the good fortune to be remarked by F. Dyson who wrote the following lines in a wonderful article about various "missed opportunities" in theoretical physics:

> «The eight groups can then be visualized as the vertices of a cube. P and G are the only kinematical groups that correspond to orthodox physical universes. But the other five groups are just as good, mathematically speaking. The most interesting of the heterodox groups are N and C. N describes a Newtonian universe with curved space-time. C describes a universe in which space is absolute, in contrast to the Galilei group G which has time absolute. The group C was discovered by Lévy-Leblond and called by him the Carroll group. In the Carroll universe, all objects have zero velocity although they may have nonzero momentum.
>
> Carroll was a pure mathematician who had already foreseen this possibility in 1871: "A slow sort of country," said the Queen, "Now, here, you see, it takes all the running you can do, to keep in the same place." But his mathematical colleagues once again missed an opportunity by failing to take him seriously.» [18]

Our paper had a career quite similar, although a bit more favourable, to the birth act of the Carroll group, in that it has known a fast increasing number of quotations after 2000 (Figure 5).

## 5 A few conclusions

— The pace of contemporary science is not necessarily "fast and furious":

> «Torniamo all'antica. Sarà un progresso.» [«Let us go back to antiquity. It will be a progress.» [19]

— Sharing knowledge may help developing it:

> « L'homme ne peut jouir de ce qu'il sait qu'autant qu'il peut le communiquer à quelqu'un [et ainsi l'enrichir]. » [«One cannot enjoy one's knowledge but by sharing it [and thereby enriching it].»] [20]

— Language should be taken seriously:

> «Le parole (…) non presentano la sola idea dell'oggetto significato, ma quando piu o quando meno immagini accessorie. Ed è pregio sommo della lingua l'aver di queste parole. Le voci scientifiche presentano la nuda e circoscritta idea di quel tale oggetto e percio si chiamano termini perche determiniano e definiscono la cosa da tutte le parti.» [«Words do not convey only the sheer idea of the object signified, but also a more or less important number of related meanings and pictures. It is the utmost value of language to be thus made of words. Most scientific terms present but the bare and limited idea of the object: they may indeed be called terms, as they determine and confine the thing.»] [21]

— It is worthwhile exploring neglected opportunities:

> «Undoubtedly, there exist many more missed opportunities to create new branches of pure mathematics out of old problems of applied science.» [18]
>
> But no less undoubtedly, reciprocally and more generally, there exist many missed opportunities to solve new problems of science out of old branches of it.

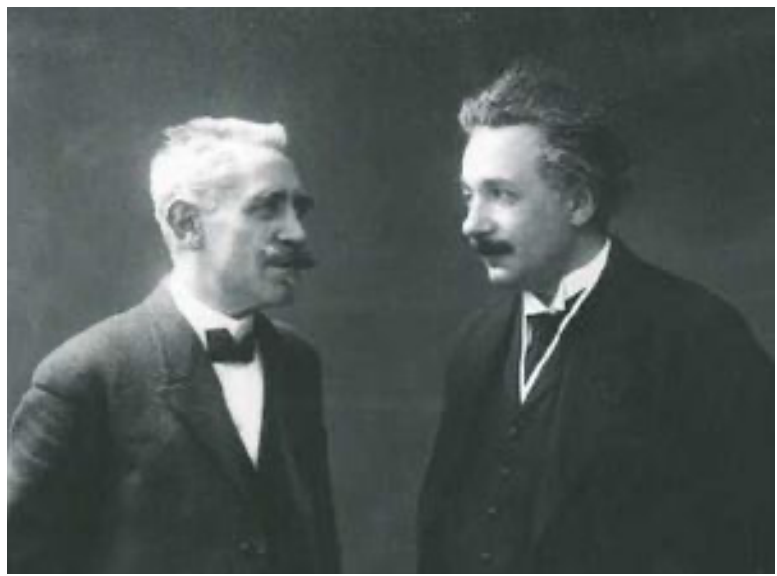

**Figure 1**
Paul Langevin and Albert Einstein in 1922

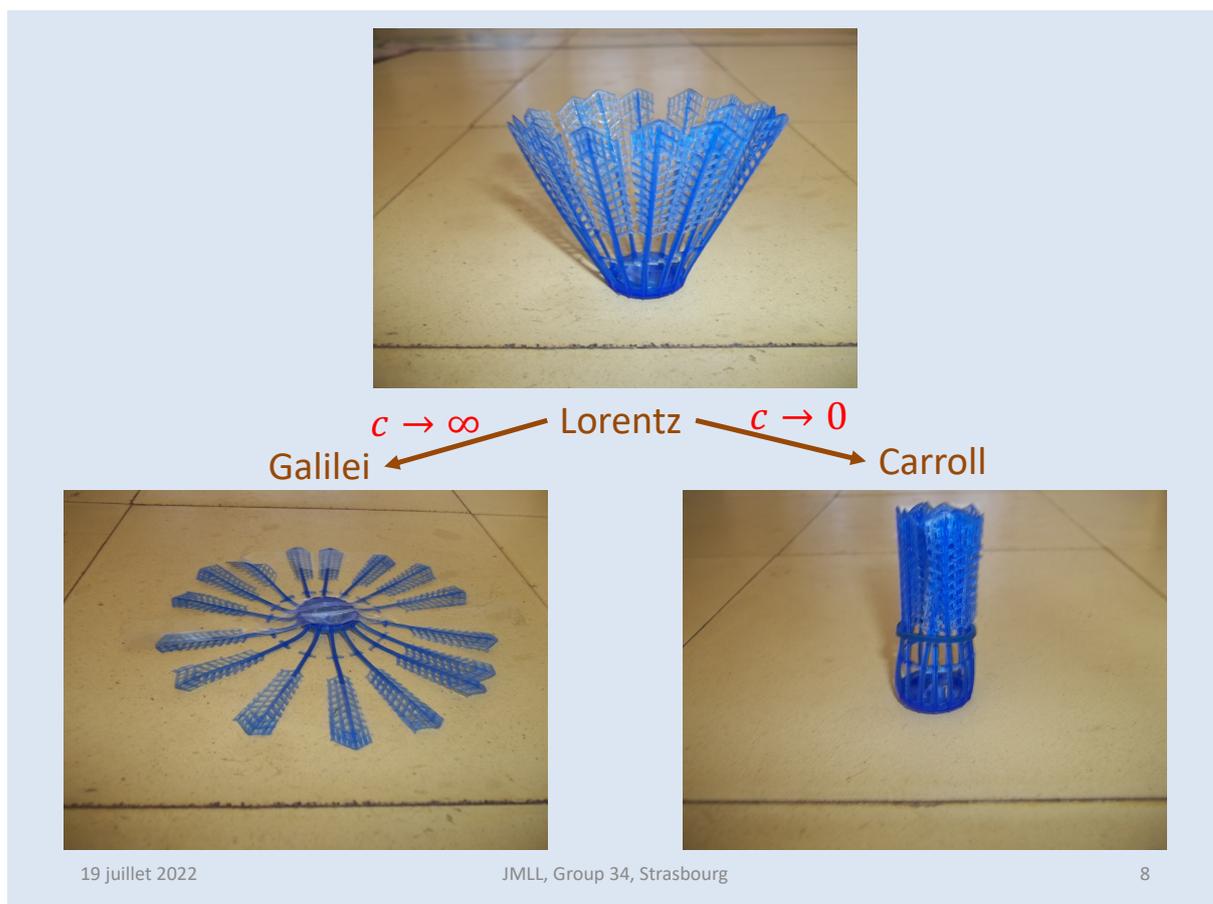

**Figure 2**

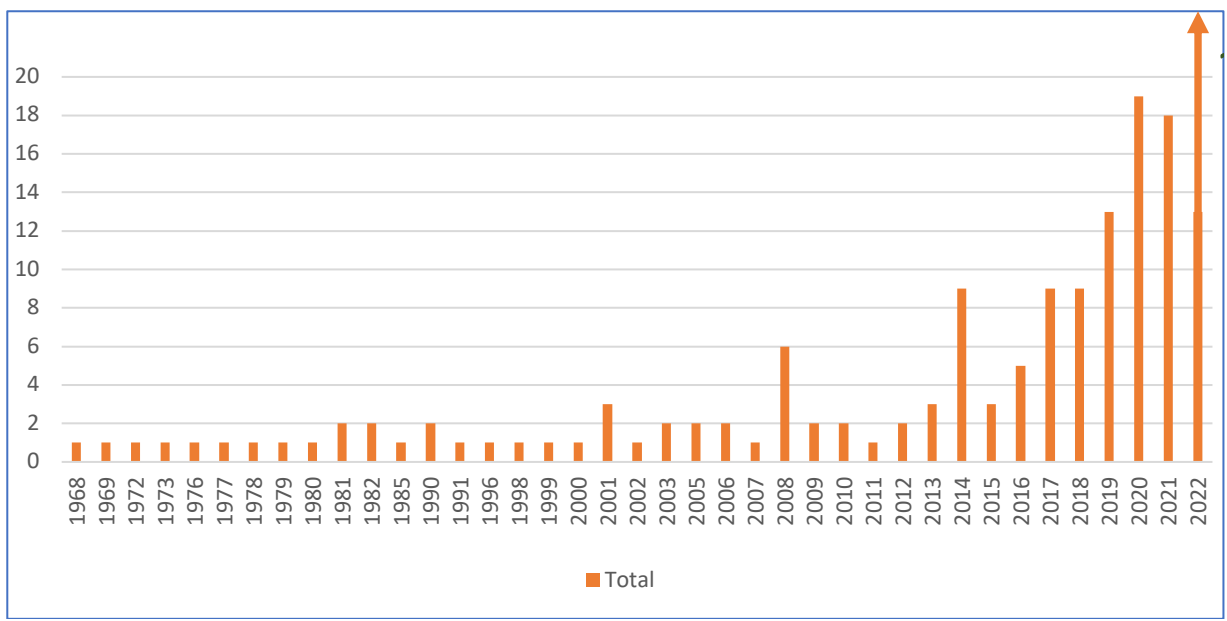

**Figure 3**
Yearly quotations of the birth certificate for the Carroll group [6]

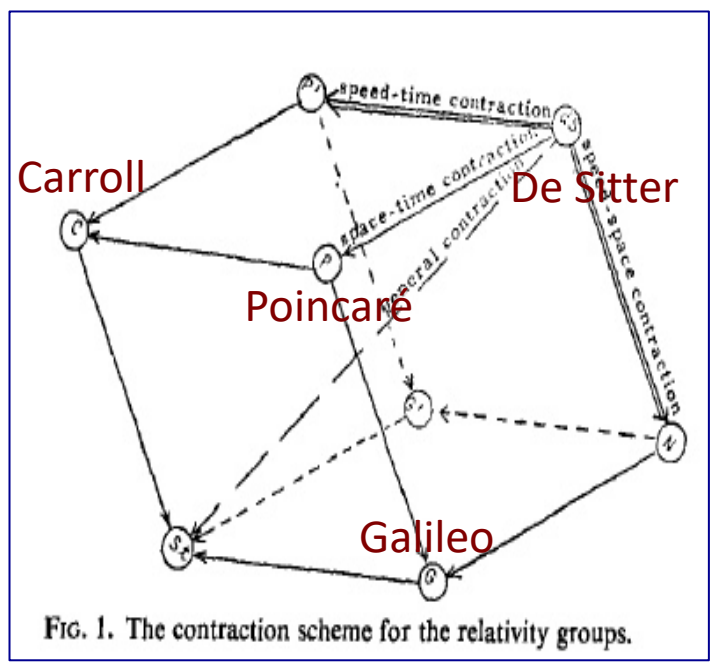

**Figure 4**
Possible chronogeometries [17]

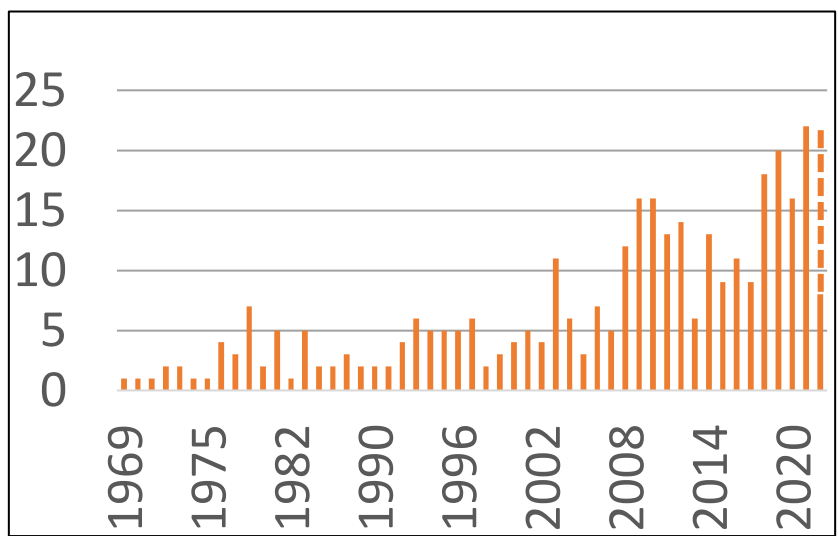

Figure 5
Yearly quotations of [17]